\documentstyle[12pt,psfig,axodraw,a4]{article}
\def\bild#1#2{    
        \vspace*{-5mm}
        \begin{center}
        \begin{math}
        \epsfxsize#2cm
        \epsffile{#1}
        \end{math}
        \end{center}
        }
\newcommand{\vs}{\vspace{-0.25cm}}

\begin{document} 
\begin{center}
\large{\bf Isospin Breaking in Neutron 
\begin{boldmath}$\beta$\end{boldmath}-decay and SU(3) Violation in
Semi-leptonic Hyperon Decays}  

\bigskip 

\bigskip

N. Kaiser\\

\bigskip

Physik Department T39, Technische Universit\"{a}t M\"{u}nchen,\\
    D-85747 Garching, Germany

\end{center}

\bigskip

\begin{abstract}
Present precision measurements of the neutron life time lead to a CKM matrix 
element $|V_{ud}|$ which is three standard deviations off the value inferred 
from heavy quark decays etc. We investigate the possibility whether 
isospin-breaking effects in the neutron-to-proton vector current  transition 
matrix element $\langle p|V^+_0|n \rangle=1+\delta g_V$ could eventually close
this gap. For that we calculate in chiral perturbation theory the effect of 
pion and kaon loops on the matrix element $\langle p|V^+_0|n \rangle$ taking 
into account the mass differences of the charged and neutral mesons. We find a
negligibly small isospin-breaking effect of $\delta g_V \simeq -4 \cdot
10^{-5}$. The crucial quantity in the analysis of neutron beta-decay precision
measurements is thus the radiative correction term $\Delta_R$. Furthermore, we
calculate in heavy baryon chiral perturbation theory the SU(3) breaking effects
on the vector transition charges of weak semi-leptonic hyperon decays. We find
for these quantities channel-dependent relative deviations from the SU(3) limit
which range from $-10\%$ to $+1\%$.
\end{abstract}


\bigskip

{\it Accepted for publication in: Physical Review C (Brief report)}

\bigskip

\bigskip

In the framework of the electroweak standard model the neutron beta-decay $n\to
p\, e^-\,\bar \nu_e$ is described by only two parameters. These are the  
quark-mixing (CKM) matrix element $V_{ud}$ and the ratio of the axial-vector
and vector coupling constants $g_A/g_V$. The (unitary) $3\times 3$ CKM-matrix 
appears already at the fundamental level of the weak interaction of quarks and
it expresses the fact that weak eigenstates and mass eigenstates of quarks are 
not identical. The ratio $g_A/g_V\ne 1$ on the other hand reflects non-trivial 
nucleon structure which has its origin in the strong interaction (i.e. QCD).  

Experimentally, both parameters can be determined from the observables in
polarized neutron beta-decay. The count rates of electrons with momentum
parallel or anti-parallel to the neutron spin define the (experimental) 
asymmetry of the electron spectrum, $A=(N^{\uparrow}-N^{\downarrow})/
(N^{\uparrow}+N^{\downarrow})$. In a recent precision measurement at ILL using 
the PERKEO II spectrometer the value  $A=-0.1189\pm 0.0008$ \cite{reich} has
been obtained for the electron asymmetry. Via the theoretical relation $A=2r(1
-r)/(1+3r^2), \,\,r=g_A/g_V$ the ratio of the axial-vector and vector coupling
constants has been deduced as $g_A/g_V = 1.2740\pm 0.0021$ \cite{reich}. The 
inverse neutron life time $\tau_n^{-1}$ on the other hand is proportional to 
$|V_{ud}|^2(g_V^2+3g_A^2)$ and therefore  it gives complementary  information. 
The proportionality factor is the product of the squared Fermi-coupling 
constant $G_F^2$ (known from muon-decay $\mu^-\to e^-\,\bar \nu_e \,\nu_\mu$), 
a three-body phase space integral depending only on the neutron-proton mass 
difference $M_n-M_p$ and the electron mass $m_e$, and a radiative correction 
term $1+\Delta_R$ to be discussed later. From the world 
average of the neutron life time $\tau_n = (885.8\pm 0.9)$\,sec (see Table\,3 
in ref.\cite{abele}) the value of the CKM matrix element $|V_{ud}|=0.9713\pm
0.0014$ \cite{abele} has been obtained using the precise empirical ratio
$g_A/g_V$ and the conserved vector current (CVC) hypothesis which implies
$g_V=1$. This value of $|V_{ud}|$ as extracted from neutron beta-decay alone, 
is marginally consistent with the one coming from an analysis of super-allowed
$0^+\to 0^+$ Fermi-transitions in nuclei which gives $|V_{ud}|= 0.9740 \pm
0.0010$ \cite{towner,pdg} (see also Table\,4 in ref.\cite{abele}). 

According to the unitarity of the CKM quark-mixing matrix in the standard 
model $|V_{ud}|$ is equal to $\sqrt{1-|V_{us}|^2-|V_{ub}|^2}$. With $|V_{us}| =
0.2196 \pm 0.0023$ from the analysis of $K_{e3}$ decays \cite{pdg} and the
information from $B$-decays, $|V_{ub}|/|V_{cb}|=0.08 \pm 0.02$ and $|V_{cb}|=
0.0395 \pm 0.0017$ \cite{abele,pdg}, one obtains this way $|V_{ud}|=\sqrt{1-|
V_{us}|^2-|V_{ub}|^2}= 0.9756\pm 0.0004$ \cite{abele}. The previously mentioned
result for $|V_{ud}|$ derived from neutron beta-decay differs from the one
inferred from unitarity by about three standard  deviations. Even though the
relative deviation of both $|V_{ud}|$-values is small (about 4 permille) it is
taken serious and considered as a possible hint at physics beyond the standard
model. In fact ref.\cite{abele} has deduced from this deviation bounds on the
masses, mass-ratio and mixing angle of new weak gauge bosons coupling to 
right-handed currents. 

However, it should be kept in mind that isospin-symmetry is not a perfect 
symmetry of the standard model and a small 4 permille deviation of the vector 
coupling constant from unity, $g_V=0.996$, would immediately resolve the 
abovementioned discrepancy. It is the purpose of the present short note to 
investigate this possibility. The result of our (exploratory) calculation is 
that isospin-breaking effects in $g_V$ are typically a factor 100 smaller. 
Therefore with respect to the accuracy of present and upcoming neutron 
beta-decay experiments one can safely use the CVC (or isospin symmetry) 
relation  $g_V=1$. 

Let us first define the V-A transition current matrix element relevant for 
neutron beta-decay,
\begin{equation} \langle p| V_\mu^+-A_\mu^+|n \rangle = \bar u_p \gamma_\mu (
g_V- g_A \, \gamma_5 ) u_n = (g_V, g_A\, \vec \sigma) \,. \end{equation}
Here, $V^+_\mu = \bar u \gamma_\mu d$ and $A^+_\mu = \bar u \gamma_\mu \gamma_5
d$ are the charge-raising vector and axial-vector currents expressed in terms
of the up- and down-quark fields. $u_{p,n}$ denote Dirac-spinors for a proton 
and a neutron (at rest) and $\vec \sigma$ is the usual Pauli spin-vector. In
the right hand side of eq.(1) one has already neglected the small four-momentum
transfer between the neutron and the proton as well as the related nucleon form
factor effects. The latter are at most of the size $(E_{max} r_N)^2/3 \simeq
1\cdot 10^{-5}$ with $E_{max}=M_n-M_p=1.293 $\,MeV the maximal energy transfer
and $r_N\simeq 0.9$\,fm a typical nucleon (electromagnetic) root mean square
radius. Of the same size is the correction from the weak magnetism $(\mu_p-
\mu_n)(E_{max}/M_p)^2$ with $\mu_p=2.793$ and $\mu_n=-1.913$ \cite{pdg} the 
proton and neutron magnetic moments.         
    
The standard model has two sources of isospin symmetry violation: the 
mass-difference of the up- and down-quarks and electromagnetic corrections. 
Prominent manifestations thereof in the hadron spectrum are the mass 
differences of the charged and neutral pions and kaons, $m_{\pi^+}-m_{\pi^0} = 
4.6$\,MeV and $m_{K^0}-m_{K^+} = 4.0$\,MeV \cite{pdg}. The $\pi^+-\pi^0$ 
mass-splitting (a sizeable $3.4\%$ effect) is almost entirely due to the 
electromagnetic interaction. A calculation of the corresponding one-photon loop
self-energy diagram employing a simple vector-meson-dominance expression
$F_\pi(t)= m_\rho^2/(m_\rho^2-t)$ for the pion charge form factor $F_\pi(t)$
gives for the $\pi^+-\pi^0$  mass-splitting   
\begin{equation} m_{\pi^+}-m_{\pi^0} = {\alpha \over 2\pi} m_\rho \Big[ x+2x^3
\ln 2x -(2x^2+1) \sqrt{x^2-1} \ln(x+\sqrt{x^2-1})\Big] = 4.3 \,{\rm MeV} \,, 
\end{equation}
with $\alpha = 1/137.036$ the fine structure constant, $m_\rho = 769\,$MeV the
(neutral) $\rho$-meson mass and $x=m_\rho/2m_{\pi^0} = 2.85$. The $K^0-K^+$ 
mass-splitting (a $0.8\%$ effect) is of different composition. Electromagnetic
effects contribute about $-2.2\,$MeV and the remaining $6.2\,$MeV are 
attributed to the  up- and down-quark mass-difference. One of the main aspects
of nucleon structure at low energies is the meson cloud surrounding the 
nucleon. In the weak $n\to p$ vector current transition neutral virtual mesons 
are converted into positively charged ones of slightly different mass and this
induces some (small) deviation of $g_V$ from unity. 

The systematic method to quantify such an isospin-breaking effect is 
chiral perturbation theory (for a review see \cite{chpt}). Observables are 
calculated with the help of an effective field theory formulated in terms of 
the Goldstone bosons ($\pi,K,\eta$) and the low-lying baryons. A systematic
expansion in small external momenta and meson masses is possible. For the 
problem considered here this means that one has to compute the quantity 
$\delta g_V$ defined by the matrix element $\langle p| V^+_0 |n\rangle = 
g_V =1+ \delta g_V$ from the one-loop diagrams shown in Fig.\,1. The relevant
effective Lagrangians, Feynman rules and loop functions can be found in
ref.\cite{chpt}. It is instructive to present first results for
$\delta g_V$ which follow from the individual pion one-loop 
diagrams (a), (b), (c), (d) and (e) shown in Fig.\,1. One finds:     
\begin{eqnarray} \delta g_V^{(a)}&=&-{g_{A}^2 m_{\pi^+}^2\over (4\pi f_\pi)^2}
\Big( 3 \ln {m_{\pi^+}\over \lambda} +1 \Big) -{g_{A}^2 m_{\pi^0}^2 \over 2(4
\pi f_\pi)^2}\Big( 3 \ln {m_{\pi^0}\over \lambda} +1 \Big) \,,\\
\delta g_V^{(b)} &=&  -{g_{A}^2 m_{\pi^0}^2 \over 2(4
\pi f_\pi)^2}\Big( 3 \ln {m_{\pi^0}\over \lambda} +1 \Big) \,,\\
\delta g_V^{(c)} &=& -{m_{\pi^+}^2 \over (4\pi f_\pi)^2}
 \ln {m_{\pi^+}\over \lambda}  -{ m_{\pi^0}^2 \over (4
\pi f_\pi)^2} \ln {m_{\pi^0}\over \lambda}  \,,\\
\delta g_V^{(d)} &=& {1 \over (4\pi f_\pi)^2(m_{\pi^+}^2-
m_{\pi^0}^2) } \bigg[ m_{\pi^+}^4 \Big(\ln {m_{\pi^+}\over \lambda} -{1\over 4}
\Big) -m_{\pi^0}^4\Big(\ln {m_{\pi^0}\over\lambda}-{1\over 4} \Big) \bigg]
\,,\\ \delta g_V^{(e)} &=& {g_{A}^2 \over (4\pi f_\pi)^2
(m_{\pi^+}^2-m_{\pi^0}^2) } \bigg[ m_{\pi^+}^4 \Big(3\ln {m_{\pi^+}\over 
\lambda}+{1\over 4}\Big) -m_{\pi^0}^4\Big(3\ln {m_{\pi^0}\over\lambda}+{1\over
4} \Big) \bigg] \,.\end{eqnarray}
Here, we have used dimensional regularization and minimal subtraction (see
appendix B in \cite{chpt}) to evaluate divergent loop integrals. The 
renormalization scale $\lambda \sim 1\,$GeV does not play a role for the final
result since the total sum of the five terms in eqs.(3-7) is in fact 
$\lambda$-independent. $f_\pi=92.4\,$MeV is the weak pion decay constant. We 
use here $g_A=1.3$ which corresponds via the  Goldberger-Treiman relation
$g_{\pi N}=g_A M_p/f_\pi$ to a strong $\pi NN$ coupling constant of $g_{\pi N}=
13.2$ which is consistent with present  empirical determinations
\cite{pavan}. Summing up the five terms given in eqs.(3-7) and expanding in the
pion mass-splitting one gets 
\begin{eqnarray} \delta g_V^{(\pi-loop)}& =& {3g_A^2+1 \over (4\pi f_\pi)^2}
\bigg\{ {m^2_{\pi^+}m^2_{\pi^0} \over m^2_{\pi^+}-m^2_{\pi^0}}\, \ln{m_{\pi^+}
\over m_{\pi^0}}-{1\over 4}(m^2_{\pi^+}+m_{\pi^0}^2) \bigg\} \nonumber \\ 
&\simeq & -\Big( g_A^2 +{1\over 3} \Big) \bigg({m_{\pi^+} - m_{\pi^0} \over 4
\pi f_\pi} \bigg)^2 = -3.2 \cdot 10^{-5}\,. \end{eqnarray}
Omitted terms of the order $(m_{\pi^+} - m_{\pi^0})^4$ are numerically 
irrelevant.  Note that eq.(8) has been derived with isospin-symmetric
interaction vertices and isospin-violating pion-propagators.  According to QCD
sum rule calculations \cite{qcdsr}  the charged and neutral pion-nucleon
coupling constants squared differ by at most 0.5\%. This is also confirmed by 
the phenomenology of charge independence breaking  in the singlet NN-scattering
lengths \cite{machleidt}. The expression in eq.(8) therefore represents indeed
the dominant isospin-breaking effect of the nucleon's pion cloud. The same set
of one-loop  diagrams (see Fig.\,1) with pions replaced by kaons gives a
further contribution of the form   
\begin{eqnarray} \delta g_V^{(K-loop)} &=& {6DF-D^2+3F^2+1\over(8\pi f_\pi)^2}
\bigg\{ {2m^2_{K^0}m^2_{K^+} \over m^2_{K^0}-m^2_{K^+}}\, \ln{m_{K^0}\over 
m_{K^+}}-{1\over 2}(m^2_{K^0}+m_{K^+}^2)\bigg\} \nonumber \\ &\simeq& -{1\over
6} ( 6DF-D^2+ 3F^2 +1) \bigg( {m_{K^0} - m_{K^+} \over 4 \pi f_\pi} \bigg)^2 =
-0.7 \cdot 10^{-5}\,. \end{eqnarray}
Here, $D\simeq 0.8$ and $F\simeq 0.5$ denote the SU(3) axial vector coupling 
constants with $g_A=D+F$. As expected the kaon cloud effect is considerably
smaller than the pion cloud effect. Further pion-loop diagrams (a), (b) and (e)
with intermediate spin-isospin-3/2 $\Delta(1232)$-excitation give rise to a
(relatively small) contribution of the form
\begin{equation} \delta g_V^{(\pi\Delta-loop)} ={g_A^2 \over \zeta^2 -1} \bigg[
{\zeta \over \sqrt{\zeta^2-1} }\, \ln\Big(\zeta + \sqrt{\zeta^2-1}\Big) -1 
\bigg] \bigg( {m_{\pi^+} - m_{\pi^0} \over 4 \pi f_\pi} \bigg)^2 = +0.4 \cdot
10^{-5}\,, \end{equation}
where $\zeta = \Delta/m_{\pi^0}=2.17$. Here, $\Delta =293\,$MeV denotes the 
delta-nucleon mass-splitting and we have used the empirically well-satisfied
coupling  constant relation $g_{\pi N\Delta}= 3g_{\pi N}/\sqrt{2}$. The fact
that $\delta g_V^{(\pi,K,\pi\Delta-loop)}$ scale with  the square of the
meson-mass splittings is consistent with the theorem of Behrends and Sirlin
\cite{behrends}. Their estimate $\delta g_V \simeq 10^{-6}$ appears however
too small in comparison to the result of our explicit calculation.

\bigskip

\begin{center}
(a) \hskip 4.cm (a) \hskip 4.cm  (b)
\end{center}
\vskip -0.7cm
\bild{gvren1.epsi}{14}
\begin{center}

\bigskip

(c) \hskip 4.cm (d) \hskip 4.cm  (e)
\end{center}
\vskip -1.0cm
\bild{gvren2.epsi}{16}
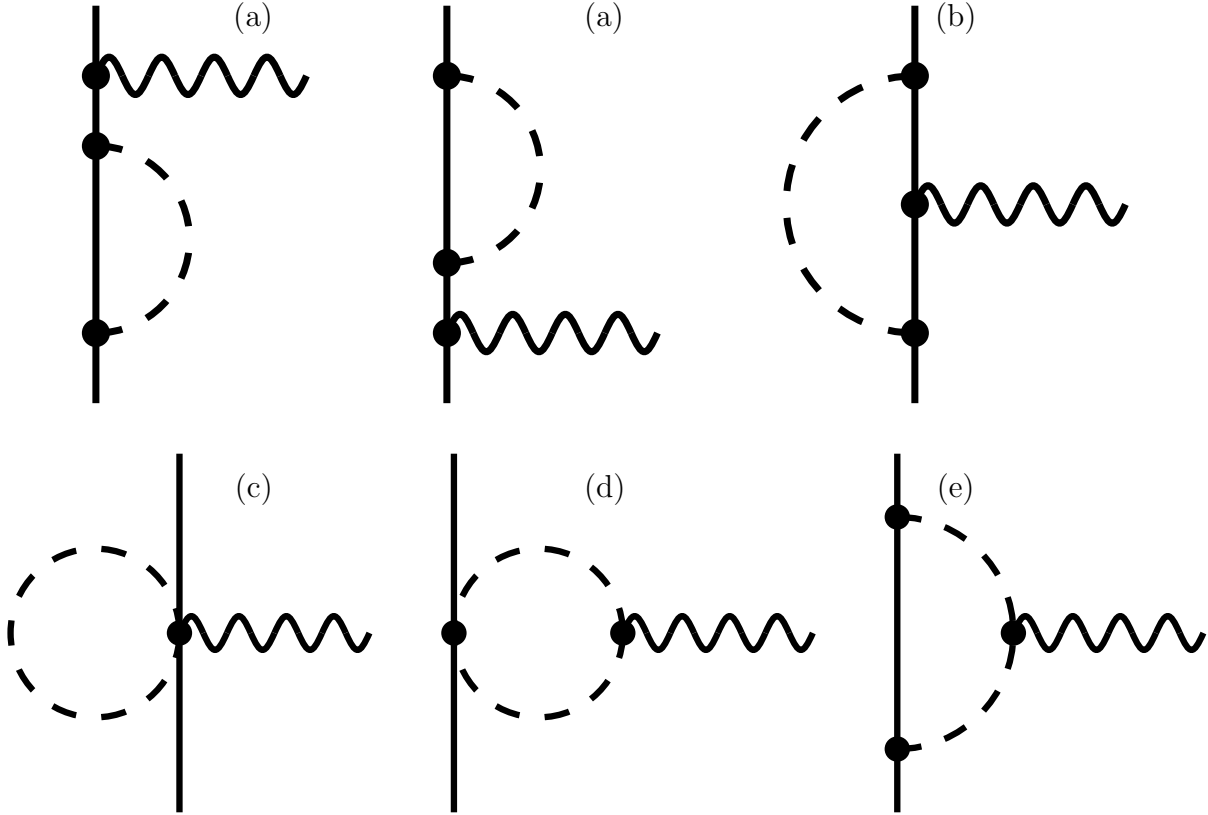
{\it Fig.\,1: One-loop diagrams for $\delta g_V$. Dashed lines represent pions,
kaons or etas. The wiggly line denotes the external charged vector field,
i.e. the $W^+$-boson.}

\bigskip

In summary, we find a negligibly small isospin-breaking effect of $\delta g_V
\simeq -4 \cdot 10^{-5}$ which does not help to resolve the presently existing 
discrepancy between various determinations of $|V_{ud}|$. Stated differently, 
we can conclude that the CVC relation $g_V^2=1$ holds with an accuracy of 
$10^{-4}$ or better. The crucial quantity in the analysis of neutron beta-decay
precision experiments is therefore the radiative correction term $\Delta_R$ 
calculated by Sirlin \cite{towner,sirlin},    
\begin{equation} \Delta_R = {\alpha \over 2\pi} \bigg( 3 \ln{M_Z\over M_p}
+\ln{M_Z\over M_A} +2C_{Born} +\dots \bigg) = (2.46 \pm 0.09) \cdot 10^{-2}\,. 
\end{equation}
Here, $M_Z= 91.19$\,GeV is the $Z^0$-mass and $M_A$ an arbitrary mass parameter
introduced in the nucleon axial form factor in order to cope with infrared
divergences. In practical applications $M_A$ is allowed to vary in the range
$0.4$\,GeV\,$<M_A<$\,$1.6$\,GeV \cite{towner,garcia}. We intend to take a 
fresh look at radiative corrections in neutron beta-decay using the systematic 
framework of chiral perturbation theory with the aim of improving on the 
quantity $\Delta_R$.   

\medskip

While isospin breaking effects in neutron beta-decay turn out to be negligibly
small, one expects much larger effects from SU(3) violation  (i.e. the mass 
difference between the strange and the up/down-quark) in strangeness-changing 
semi-leptonic weak hyperon decays. Considering the matrix-elements of the 
charged strangeness-changing vector current $\bar u\gamma_\mu s$ at momentum  
transfer zero, the Ademollo-Gatto theorem \cite{gatto} asserts however that 
for these quantities SU(3) breaking effects start first at quadratic order in 
the quark mass difference $m_s-m_{u,d}$. In order to quantify these SU(3)
breaking effects we calculate here in heavy baryon chiral perturbation theory
the leading order contributions arising from $(\pi,K,\eta)$-loop diagrams 
(see Fig.\,1). In such a calculation SU(3) symmetry breaking originates 
entirely from the different masses of the eight pseudoscalar Goldstone bosons
$(\pi,K,\eta)$. The quantities of interest are the matrix elements of the
strangeness-changing ($s\to u)$ weak  vector current density evaluated in
baryon states which differ by one unit of strangeness, 
\begin{equation} \langle B'| \bar u \gamma_0 s| B\rangle = g_V(B\!\to\!B')\Big[
1+\delta_V(B\!\to\! B') \Big]\,, \end{equation}
with $B\in\{ \Lambda,\Sigma^0,\Sigma^-,\Xi^0,\Xi^-\}$ and $B'\in\{
p,n,\Lambda,\Sigma^+,\Sigma^0\}$. The numbers $g_V(B\!\to\!B')$ correspond to 
the results in the exact SU(3)-limit and they read,
\begin{eqnarray} && g_V(\Lambda\!\to\! p) = -{\sqrt 6 \over 2}\,, \qquad  
g_V(\Sigma^0\!\to\! p)=-{\sqrt 2 \over 2}\,,\qquad  g_V(\Sigma^-\!\to\!n)=-1\,,
\nonumber \\ && g_V(\Xi^0\!\to\!\Sigma^+)=1\,, \qquad g_V(\Xi^-\!\to\! \Lambda)
={\sqrt 6 \over 2}\,, \qquad g_V(\Xi^-\!\to\!\Sigma^0)={\sqrt 2 \over 2}\,. 
\end{eqnarray}
The quantities $\delta_V(B\!\to\! B')$ measure for a specific transition the 
relative deviation from SU(3) symmetry. Evaluation of the one-loop diagrams
shown in Fig.\,1 leads to the following expression for the SU(3) breaking
effect, 
\begin{eqnarray} \delta_V(B\!\to\! B') &=& {1\over (8\pi f_\pi)^2} \bigg\{
\Big[3+\alpha(B\!\to\! B')\Big] \bigg[ {m_K^2 m_\pi^2 \over m_K^2 -m_\pi^2} 
\, \ln {m_K \over m_\pi} -{1\over 4}( m_K^2 +m_\pi^2) \bigg] \nonumber \\ && +
\Big[3+\beta(B\!\to\!B')\Big] \bigg[{m_\eta^2 m_K^2 \over m_\eta^2 -m_K^2} \, 
\ln {m_\eta \over m_K} -{1\over 4}( m_\eta^2 +m_K^2) \bigg] \bigg\} \,,
\end{eqnarray} 
with the channel-dependent coefficients $\alpha(B\!\to\! B')$ and
$\beta(B\!\to\! B')$ given by 
\begin{equation} \alpha(\Lambda \!\to\! p ) = 9D^2+6DF+9F^2\,, \qquad
\beta(\Lambda \!\to\! p ) = (D+3F)^2\,, \end{equation} 
\begin{equation}\alpha(\Sigma^0\!\to\! p)=\alpha(\Sigma^-\!\to\! n) =
D^2-18DF+9F^2\,, \qquad \beta(\Sigma^0\!\to\! p)=\beta(\Sigma^-\!\to\!
n)=9(D-F)^2\,,\end{equation}  
\begin{equation} \alpha(\Xi^-\!\to\! \Lambda)= 9D^2-6DF+9F^2\,, \qquad 
\beta(\Xi^- \!\to\!\Lambda )=(D-3F)^2\,,\end{equation} 
\begin{equation}\alpha(\Xi^0\!\to\! \Sigma^+ )=\alpha(\Xi^-\!\to\!\Sigma^0 ) 
= D^2+18DF+9F^2\,, \qquad \beta(\Xi^0\!\to\!\Sigma^+) =\beta(\Xi^-\!\to\! 
\Sigma^0)=  9(D+F)^2\,. \end{equation} 
The channel-independent terms in eq.(14) proportional to the coefficient 3 
stem from the ($\pi,K,\eta)$-loop diagrams c) and d) in Fig.\,1 with no 
internal heavy baryon propagator. For the numerical evaluation of
$\delta_V(B\!\to \!B')$ we use for the axial-vector coupling constants $D=
0.8$, $F=0.5$ and for the meson masses $m_\pi=139.57\,$MeV, $m_K= 493.68\,$MeV,
$m_\eta=\sqrt{(4m_K^2-m_\pi^2)/3}=564.33\,$MeV (the value from the GMO-relation
which deviates only by 3.1\% from the physical $\eta$-mass). This input gives
numerically,   
\begin{equation}\delta_V(\Lambda\!\to\! p)\simeq -10\%\,, \qquad \delta_V(
\Sigma^0\!\to\! p)=\delta_V (\Sigma^-\!\to\! n)\simeq +1\%\,, \end{equation} 
\begin{equation} \delta_V(\Xi^-\!\to\! \Lambda)\simeq -6\%\,, \qquad 
\delta_V(\Xi^0\!\to\! \Sigma^+)=\delta_V(\Xi^-\!\to\! \Sigma^0) \simeq -10\%
\,. \end{equation}  
One notices that the deviations from SU(3) symmetry are sizeable and strongly
channel-dependent (for results of a relativistic calculation, see 
ref.\cite{krause}). Clearly, these SU(3) breaking effects in the weak vector 
transition matrix elements should be included in the analysis of the 
strangeness-changing semi-leptonic hyperon decays. On the theoretical side one 
should attempt to further improve the predictions for $\delta_V(B\!\to\! B')$
by performing  next-to-leading (or even higher) order calculations in baryon
chiral  perturbation theory. 

Let us finally make a comparison to SU(3) breaking
effects in the axial vector coupling constants $D$ and $F$. Chiral logarithmic
corrections of the form $(m_K/4\pi f_\pi)^2 \ln(m_K/\lambda)$ to $D$ and $F$ 
have been calculated in ref.\cite{jenkins}. With inclusion of only octet 
baryons in intermediate states the tree-level couplings $D$ and $F$ had to be 
reduced by about $30\%$ and the further inclusion of decuplet intermediate  
states increased these couplings again somewhat. The chiral corrections to the 
vector couplings $\delta_V(B\to B')$ considered here behave differently. First,
the one-loop results are finite (i.e. independent of the renormalization scale
$\lambda$), free of possible counterterm contributions and secondly the 
relative deviations from the SU(3)-limit do not
exceed $10\%$. Of course only complete higher order calculations allow to judge
the accuracy of the present one-loop results. In order test the SU(3) breaking
effects for the vector couplings $\delta_V(B\to B')$ a reanalysis of
strangeness-changing semi-leptonic hyperon decays \cite{bourquin} would be most
useful.

\end{document}